\documentclass{article}
\usepackage[dvips]{graphicx}
\begin{document}
\title{Folksonomy as a Complex Network}
\author{\textbf{Kaikai Shen, Lide Wu}\\ Department of Computer Science\\Fudan University\\Shanghai, 200433 \thanks{Supported by \emph{Chun-Tsung Undergraduate Research Endowment}}}

\date{}\maketitle
\begin{abstract}
Folksonomy is an emerging technology that works to classify the
information over WWW through \emph{tagging} the bookmarks, photos or
other web-based contents. It is understood to be organized by every
user while not limited to the authors of the contents and the
professional editors. This study surveyed the folksonomy as a
complex network. The result indicates that the network, which is
composed of the tags from the folksonomy, displays both properties
of \emph{small world} and \emph{scale free}. However, the statistics
only shows a local and static slice of the vast body of folksonomy
which is still evolving.

\end{abstract}

\section{Introduction}

\subsection{Folksonomy and Tags}

The etymology of the word \textit{Folksonomy} shows that it's a
portmanteau of the words \textit{folks} and \textit{taxonomy} coined
by Thomas Vander Wal\cite{atom}, which implies that it could be
understood as an organization  by folks, especially of the contents
over the world wide web. Being different from the traditional
approaches to the classification, the classifiers in folksonomy are
not the dedicated professionals, and Thomas Vander Wal described
this as a "bottom-up social classification"\cite{wal}. Adam Mathes
explains folksonomy that users of the documents and media create
metadata - data about data - for their own individual use that is
also shared throughout a community\cite{mathes}.

Del.icio.us (\verb"http://del.icio.us"), Furl
(\verb"http://www.furl.net") and Flickr
(\verb"http://www.flickr.com") are three most popular folksonomies.
Their users describe and organize the content (bookmarks, webpages
or photos) with their own vocabulary and assign one or more
keywords, namely \emph{tags}, to each single unit of content. The
folksonomy is thus implemented through the tags assigned. Therefore,
\emph{tags} are now the mainstream approach to the application of
folksonomy, and folksonomy is currently often understood as
\emph{tagging}.

\subsection{Folksonomy as Network}
As was mentioned above, folksonomy enables users to share their
individual use of tags in the community. Users share various
contents under one same tag, or share different tags assigned to one
piece of content. Thus tags are linked to each other and so are the
contents. Such a feature makes it possible to understand the
folksonomy a network of tags or contents.

Besides the network of folksonomy, some similar networks were
reported to display the properties of small world or scale
free.\cite{albert} It is possible to measure the graph properties of
 World Wide Web in order to quantify the information therein and give out an the
 explanation or its evolution.\cite{DHYANI} In 2001 Ferrer i Cancho and Sol\'{e} defined a network in English
language. Another study by Yook, Jeong and Barab\'{a}si constructed
a network based on the synonym according to Merriam-Webster
Dictionary. They observed a small average path length clustering
coefficient and power-law degree
distribution\cite{ferrer}\cite{albert}, and indicated that language
also forms a complex network in some respects. Rosa Gil \textit{et
al.}\cite{gil} model and analyze the semantic web as complex system.

In the light of these works and results, the network of folksonomy
can be defined and constructed. While comparing this network with
that modeled by Rosa Gil \textit{et al.}\cite{gil}, the difference
lied mainly in the difference between the \emph{tags} and the
ontologies in the \emph{DAML Ontology Library}.

\section{Properties of Folksonomy Network}
In order to learn the conformation of the folksonomy network
realized through tags, to see whether it displays such properties of
small world or scale free and to measure the folksonomy, the model
of the network must be defined first. Folksonomy can be considered
as a graph where nodes represent the tags and different tags
assigned to one piece of content are linked by edges. This graph is
an undirected graph. Regardless of multiple contents covering two
tags, the graph is not a weighted graph.
\begin{description}
    \item[Degree distribution] For an selected node $i$ in the folksonomy network,
    its degree $k_i$ represents the number of tags which share at
    least one piece of content with the tag (or node) $i$. For each
    network, the spread in node degree follows a distribution
    function $P(k)$. For \emph{scale free} networks, the degree is
    in a power-law distribution
    $$P(k)\sim k^{-\gamma}.$$
    \item[Clustering coefficient] For node $i$ with the degree
    $k_i$, it is connected with $k_i$ nodes in the network. There
    are $E_i$ edges in this subgraph of $k_i$ size, and could be
    at most $C_{k_i}^2=\frac{1}{2}k_i(k_i-1)$ edges between these
    $k_i$ nodes. The ratio
    $$C_i=\frac{2E_i}{k_i(k_i-1)}$$is the clustering
    coefficient of the node $i$. The clustering coefficient $C_i$ measures
    the interrelatedness of $i$'s neighbors.
    \item[Average path length] For two nodes $i, j$ in the same connected
    component, $l_{ij}$ is the minimum length of path between them. The
    average path length $l$ is the average value of all $l_{ij}$.
\end{description}
\section{Experiment}
The data set of the experiment is based on the records of the
bookmarks submitted to Del.icio.us during 26 Mar. to 27 Mar., 2005.
Del.icio.us provides the service that enables users to categorize
their bookmarks or links with \emph{tags}.

All the data used in this experiment is available through the
subscription of RSS feed of Del.icio.us
(\verb"http://del.icio.us/rss"). For each entry of the bookmarks,
only the information of the URL, the time of submission and the tags
were recorded. Other information as the creator, the title was
ignored in the experiment.

For every distinct URL, all the tags attributed to it will be linked
to each other with edges. The network is thus constructed.

\subsection{Folksonomy as a Small World Network}
Random networks were first defined by P. Erd\"{o}s and A. R\'{e}nyi
in 1959. In such a \emph{random network} of Erd\"{o}s-R\'{e}nyi
model, the average path length $l_{random}$ is small with regard to
the size $N$ of the network,
$$l_{random}\simeq\frac{\ln{N}}{\ln{\langle k\rangle}}$$ and its clustering
coefficient
$$C_{random}\simeq\frac{\langle k\rangle}{N}.$$

The \emph{small world} network of Watt-Strogatz
displays\cite{Strogatz}\cite{watts}, as the random network with the
same $N$ and $\langle k\rangle$, the similar property of small
average path length $$l\simeq l_{random}$$however with a relatively
high clustering coefficient$$C\gg C_{random}.$$

The properties of the network of folksonomy tags in experiment turns
out as follows.

\begin{itemize}
    \item \textbf{Nodes (the number of tags)} $N$:  9804
    \item \textbf{Average node degree} $\langle k\rangle$:  11.0
    \item \textbf{Clustering coefficient} $C$:  0.06
    \item \textbf{Average path length} $l$:  3.40
\end{itemize}

For the network in the experiment, its average path length $l=3.40$
is approximately the length $l_{random}\simeq 3.83$ of the
corresponding random network. And its clustering coefficient
$C=0.06$ is much larger than the prediction $C_{random} \simeq
0.001$ if the network is random. Therefore It can be concluded to be
an small world network.

\subsection{Folksonomy as a Scale Free Network}
Lots of real networks are reported to be \emph{scale
free}\cite{albert}, i.e. its degree distribution $P(k)$ is in
power-law
$$P(k)\sim k^{-\gamma}.$$
While in Erd\"{o}s and R\'{e}nyi's theory, the degree distribution
$P(k)$ of a random network will follow Poisson distribution.

The property of scale free can be detected in the folksonomy
network. Figure 1 indicates the distribution is linear in
logarithmic scale, as well as its Complementary Cumulative
Distribution Function, CCDF, in Figure 2. The result from the
folksonomy network (see Figure 1) shows its degree distribution
decays at the rate of $k^{-\gamma}$, where the power-law exponent
$\gamma$ is $1.418$.

Table 1 is a top-20 list of tags involved in experiment with the
most degree in the network, namely, those have the most contacts
with the other tags.

\begin{table}{\centering
\begin{tabular}{r|p{2cm}}\hline
\textsc{degree} & \textsc{tag's name}\tabularnewline
  \hline
  1504  &  blog\tabularnewline 1465  &  web\tabularnewline 1297&
software\tabularnewline 895& music\tabularnewline 724&
design\tabularnewline 631& art\tabularnewline 467&
programming\tabularnewline 326& reference\tabularnewline 265&
tools\tabularnewline 205& news\tabularnewline 188&
cool\tabularnewline 163& linux\tabularnewline 162&
mac\tabularnewline 115& internet\tabularnewline 111&
howto\tabularnewline 108& blogs\tabularnewline 94&
technology\tabularnewline 86& fun\tabularnewline 81&
science\tabularnewline 47& tech\tabularnewline \hline
\end{tabular}
\caption{Top 20 degree tags}}
\end{table}

\section{Conclusion and Future Work}

The experiment samples a part of the folksonomy at Del.icio.us,
which demonstrated above that the folksonomy as a network formed by
tags displays both nature of \emph{small world} and \emph{scale
free}.

However the folksonomy network is said to be \emph{small world} and
\emph{scale free} as local properties. The body of folksonomy is
much larger than this fragment. All tags over WWW indexed by
Technorati are more than 1 million\cite{technorati}. It is possible
that the panorama of folksonomy and the parameters of the whole
network would differ from the present local ones.

Since users and authors over WWW submits their contents to
folksonomy every minute, the network of folksonomy evolves over
time. The work in the experiment surveyed the static properties of a
folksonomy network, but the network is dynamically increasing every
moment. The study of dynamics on the complex networks will be
applied to the further analysis of folksonomy's structure, behavior
including its forming mechanism. Since folksonomy is a
classification system of web contents, its properties both static
and dynamic can also serve to search and retrieve the related
information.

\begin{figure}
{\centering
\includegraphics[width=11cm]{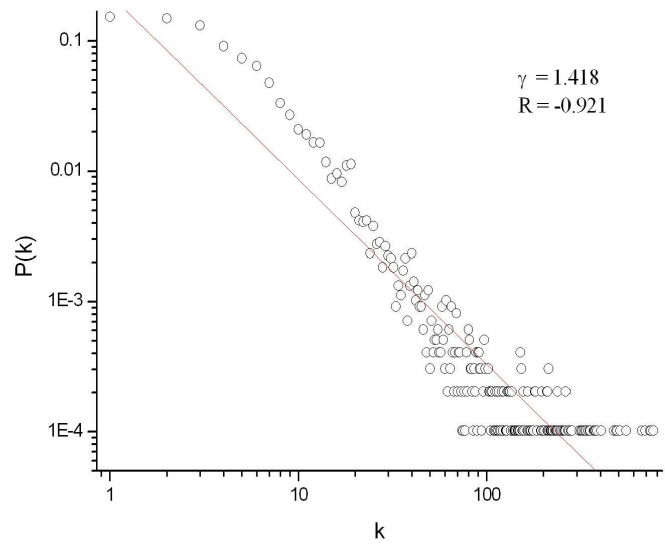}
\caption{Degree distribution of folksonomy network. In logarithmic
scale. $R$ is the correlation coefficient.} }
\end{figure}

\begin{figure}
{\centering
\includegraphics[width=11cm]{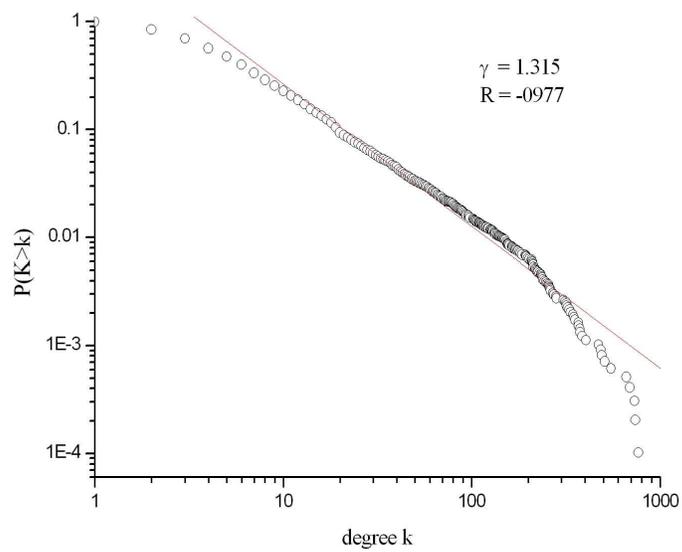}
\caption{Degree Complementary Cumulative Distribution Function,
CCDF. In logarithmic scale. $R$ is the correlation coefficient.} }
\end{figure}

\end{document}